\documentclass[12pt]{article}
\usepackage{amsmath}
\usepackage{psfig,epsfig}
\usepackage{latexsym}
\usepackage{pifont}

\newcommand{\lapprox}{\lower.6ex\hbox{$\; \buildrel<\over\sim \;$}}

\newcommand{\gapprox}{\lower.6ex\hbox{$\; \buildrel>\over\sim \;$}}

\newcommand{\curly}{\lower1.ex\hbox{$\; \stackrel{\textstyle \wr}{\wr} \;$}}

\def\barr{\begin{array}}
\def\earr{\end{array}}
\def\berr{\begin{eqnarray}}
\def\err{\end{eqnarray}}
\def\berrno{\begin{eqnarray*}}
\def\errno{\end{eqnarray*}}
\def\be{\begin{equation}}
\def\ee{\end{equation}}

\def\apj{{\it Astrophys.~J.}}

\def\prd{{\it Phys.~Rev.~D.}}
\def\prl{{\it Phys.~Rev.~Lett.}}
\def\mnras{{\it Mon. Not. R.~Astr.~Soc.}}

\def\AnA{{\it Astr. Astrophys.}}
\def\grg{{\it Gen. Rel. Grav.}}
\def\pl{{\it Phy. Lett.}}

\title{\bf Variable Chaplygin Gas: Constraints from CMBR and SNe Ia}
\author{Geetanjali Sethi{}{\footnote{getsethi@physics.du.ac.in}} ,
  Sushil K. Singh, Pranav Kumar\\ {\it Department of Physics \&
    Astrophysics,}\\{\it University of Delhi, Delhi 110 007}\\ \\Deepak Jain\\{\it
    Deen Dayal Upadhyaya College, University of Delhi, Delhi 110 015}
\\  \\ and\\ \\  Abha Dev\\{\it  Miranda
  House, University of Delhi, Delhi 110 007}}
\begin{document}

\maketitle

\begin{abstract}

We constrain the  parameters of the  variable Chaplygin gas
model, using the location of peaks of the
CMBR spectrum and SNe Ia ``gold '' data set.
Equation of state  of the model is
$P=-A(a)/\rho$, where $A(a)=A_0 a^{-n}$ is a positive function of the
cosmological scale factor $a$, $A_0$ and $n$ being constants. The
variable Chaplygin gas interpolates from dust-dominated era  to
quintessence dominated era. The model is found to be compatible  with
current type Ia Supernovae data and location of first peak if the
values of $\Omega_m$ and $n$ lie in the interval $[0.017,~0.117]$ and $[-1.3,~2.6]$ respectively.
\indent\hspace{1.2cm}- SNeIa
\end{abstract}

\section{Introduction}

It is now well established that the expansion of the
universe is accelerating. The direct evidence for acceleration
 comes from the Hubble diagram of
Type Ia Supernovae (SN Ia) \cite{perl}.
The observational results of  SN Ia  together  with
the anisotropy of cosmic microwave background radiation (CMBR) power
spectrum  and
clustering estimates show that our universe is mainly made up of two
components: dark matter and dark energy. The nature of dark matter and
dark energy is not well
understood. The dark matter contribute one-third of the total energy
density of the universe. The dark energy, which is self interacting,
unclustered fluid with large negative pressure, contributes roughly
two-third of the total energy density of the
universe(for latest review see \cite{varun}).

The simplest and the most favored candidate for the dark energy is
the cosmological constant ($\Lambda $). Consequently, several models
with a relic cosmological constant $(\Lambda CDM)$ have been used to
describe the observed universe. However most of them suffer
from severe fine tuning problem \cite{varun,weinberg}.
There are also models of dark energy with varying
equation of state \cite{hkj}.
Another possibility for dark energy is quintessence which involve a
slowly evolving and spatially homogeneous scalar field $\phi$,
\cite{ratra,zlatev} or two coupled fields \cite{santos}. However ,
quintessence models also suffer from  fine-tuning problem. This
problem is usually highlighted as the ``why now'', that is, why does
the dark energy start dominating over the matter content of the Universe recently.

As an alternative to both the cosmological
constant and quintessence,  It is also possible to explain the
acceleration of the universe by introducing a cosmic fluid component
with an exotic equation of state, called Chaplygin gas
\cite{kamen,billic}. The attractive feature
of such models are that they can explain both dark energy and dark
matter with a single component. The equation of state for the
Chaplygin gas is $P = -A/\rho$, where $A$ is a positive constant. A more
{\it generalized} model of Chaplygin gas is characterized by an
equation of state

\be
P_{ch} = -\frac{A}{\rho_{ch}^{\alpha}}
\end{equation}
where $\alpha$ is a constant in the range $0<\alpha \le 1$ (the
Chaplygin gas corresponds to $\alpha = 1$). By inserting eq.(1) into
the energy conservation law we get the expression for the energy
density as \cite{bento}
\be
\rho_{ch} = (A + \frac{B}{a^{3(1+\alpha)}})^{\frac{1}{1+\alpha}}
\end{equation}
where $a$ is the scale factor and $B$ is a constant of integration.
As can be seen from the above equation, the Chaplygin gas behaves
like a non relativistic matter at early times while at late times the
equation of state is dominated
by a cosmological constant $8\pi GA^{1/(1+\alpha)}$ which thus
leads to the observed accelerated expansion.
The Chaplygin gas models have been found to be consistent with SNeIa
data \cite{makler}, CMB peak locations \cite{bento1} and other
observational tests like gravitational lensing, cosmic age of old high
redshift objects etc.\cite{dev} , as also with combination of some of
them \cite{bean}.
It has been shown that this model can be accommodated within the
standard structure formation scenarios
\cite{billic,bento,fabris}. Therefore the Chaplygin gas model seems to
be a good alternative to explain the accelerated expansion of the
universe.However the Chaplygin gas model produces oscillations or exponential blowup of matter power spectrum that are inconsistent with observations\cite{teg}.

Recently a {\it variable} Chaplygin gas model was proposed \cite{cgas} and
constrained using SNeIa ``gold'' data \cite{reiss}. In this letter, we try to
constrain the variable Chaplygin gas model parameters using location
of CMB peaks and SNeIa observations. We give the basic formalism of the
model in section 2. Section 3 gives general formulae for CMB peaks.
The best fit parameters have been calculated in section 4 by reduced $\chi^2$ minimization
of the distance modulus, followed by discussion and conclusions in section 5.


\section{Variable Chaplygin gas model}

We consider the proposed variable Chaplygin gas \cite{cgas}
characterized by the equation of state:
\be\label{pch}
P_{ch} = -\frac{A(a)}{\rho_{ch}}
\ee
where $A(a) = A_0 a^{-n}$,  is a positive function of the cosmological
scale factor $a$. $A_0$ and $n$ are constants. Using the energy
conservation equation in flat  FRW universe and eq.\ref{pch}, the
variable Chaplygin gas density evolves as:

\be
\rho_{ch}(a) = \sqrt{\frac{6}{6-n}\frac{A_0}{a^n}+\frac{B}{a^6}}
\ee
where $B$ is a constant of integration.

\noindent The original Chaplygin gas scenario is restored for $n = 0$ and the
gas behaves initially as dust-like
matter and later as a cosmological constant. However, in the present
case the Chaplygin gas evolves from dust dominated epoch  to
quintessence in present times (for details see \cite{cgas} and
references therein).
\\
The Friedmann equation for a spatially flat universe reads as
\be\label{hubble}
H^2 = \frac{8\pi G}{3}\rho
\ee
where $H \equiv \dot{a}/a $ is the Hubble parameter. Therefore, the
acceleration condition $\ddot a > 0 $ is equivalent to

\be
\Big(3-\frac{6}{6-n}\Big)a^{6-n} > \frac{B}{A_0},
\ee
which requires $n~<~4$. This gives the present value of energy
density of the variable Chaplygin gas

\be
\rho _{ch0} = \sqrt{\frac{6}{6-n}A_0+B}
\ee
where $a_0~=~1$. Defining

\be
\Omega_m \equiv \frac{B}{6A_0/(6-n)+B},
\ee
the energy density becomes

\be\label{rho1}
\rho_{ch}(a) = \rho_{ch0}\Big[\frac{\Omega_m}{a^6} +
  \frac{(1-\Omega_m)}{a^n}\Big]^{1/2}
\ee


\section{Location of CMBR peaks}


 The location of peaks are very sensitive to the variations in the
 parameters of the model and hence serve as a sensitive probe to
 constrain the cosmological parameters and discriminate among various
 models \cite{fukugita,doran}. The peak locations are set by the
 acoustic scale $l_{A}$, which can be interpreted as angle subtended
 by the sound horizon at the last scattering surface. This angle (say
 $\theta_{A}$) is given by ratio of sound horizon to angular diameter
 distance, $d_{A}$ of the last scattering surface:

\be
\theta_{A} = \frac{a(t_{ls})\int_{0}^{t_{ls}} c_{s} \frac{dt}{a(t)}}{d_{A}(t_{ls})}
\end{equation}
where $c_{s}$ is the speed of sound given in plasma by
$c_{s}~=~1/(3(1+R))^{1/2}$ and $R~=~3 \rho_{b}/ \rho_{\gamma}$
corresponds to the ratio of baryon to photon density. For the 
flat ($k~=~0$) FRW model, the Acoustic scale
$l_{A}~=~ \pi/\theta_{A}$ is

\be\label{peak}
l_A = \pi \frac{\tau_{0}-\tau_{ls}}{c_s\tau_{ls}}
\end{equation}
where $\tau_{0}$ and $\tau_{ls}$ are conformal time today and at the last scattering surface.

For our calculations we assume: present
value of scale factor $a_{0}~=~1$, the value of scale factor at the
last scattering surface $a_{ls}~=~1/1100$, $h~=~0.71$, the density
parameter for radiation and baryons at present
$\Omega_{\gamma_{0}}~=~9.89 \times 10^{-5}$, $\Omega_{b0}~=~0.05$,
average sound speed $\bar{c_{s}}~=~0.52$ and spectral index for initial
energy density perturbations, $n~=~1$.

The location of the $i^{th}$ peak in the spectrum is given by

\be
l_{i} = l_{A}(i-\delta_i)
\end{equation}
where the phase shift $\delta_i$, caused by the plasma driving effect,
is solely determined by the pre-recombination physics. We can
approximate it with its value from standard cosmology \cite{fukugita}:

\be \label{delta}
\delta_1 \approx 0.267[\frac{r(z_{ls})}{0.3}]^{0.1}
\end{equation}
where $r(z_{ls})~=~ \rho_r(z_{ls})/\rho_m(z_{ls})~=~
\Omega_{r0}(1+z_{ls})/\Omega_{m0}$ is the ratio of radiation and matter
density at the decoupling epoch.
For the third peak,

\be
\delta_3 \approx 0.35\Big[\frac{r(z_{ls})}{0.3}\Big]^{0.1}
\end{equation}


The Friedmann equation, using Eqn. \ref{rho1} for a variable Chaplygin
gas becomes

\be\label{FRW}
H^2 = \frac{8\pi G}{3} \Big[\rho_{r0}(1+z)^4 +
  \rho_{b0}(1+z)^3 + \rho_{ch0} \Big[ \Omega_m(1+z)^6 +
    (1-\Omega_m)(1+z)^n \Big]^{1/2} \Big]
\ee
where $\rho_{r0}$ and $\rho_{b0}$ are the present values of energy
densities of radiation and baryons respectively.

Using

\be
\frac{\rho_{r0}}{\rho_{ch0}} = \frac{\Omega_{r0}}{\Omega_{ch0}} =
\frac{\Omega_{r0}}{1-\Omega_{r0}-\Omega_{b0}}
\ee
and
\be
\frac{\rho_{b0}}{\rho_{ch0}} = \frac{\Omega_{b0}}{\Omega_{ch0}} =
\frac{\Omega_{b0}}{1-\Omega_{r0}-\Omega_{b0}}
\ee
Eqn. \ref{FRW} becomes

\be
H^2 = \Omega_{ch0}H_0^2a^{-4}X^2(a)
\ee
where
\be
X^2(a) = \frac{\Omega_{r0}}{1-\Omega_{r0}-\Omega_{b0}} +
\frac{\Omega_{b0}a}{1-\Omega_{r0}-\Omega_{b0}} +
a^4\Big(\frac{\Omega_m}{a^6} + \frac{1-\Omega_m}{a^n}\Big)^{1/2}.
\ee
Using the fact that $H^2~=~a^{-4}(\frac{da}{d\tau})^2$, we get

\be
d\tau = \frac{da}{\Omega^{1/2}_{ch0}H_0X(a)},
\ee
so that Eqn.\ref{peak} becomes

\be
\it{l_{A}} = \frac{\pi}{c_s}\Big[ \int_0^1\frac{da}{X(a)}\Big(
  \int_0^{a_{ls}} \frac{da}{X(a)}\Big)^{-1} - 1\Big].
\ee
The ratio of radiation to matter density at the last scattering
surface, $r_{ls}$ in Eqn.\ref{delta} becomes

\be
r_{ls} = \frac{\Omega_{r0}}{a^4_{ls}\Big(1 - \Omega_{r0} -
  \Omega_{b0}\Big)\Big(\frac{\Omega_m}{a^6} + \frac{(1 -
    \Omega_m)}{a^n}\Big)^{1/2}}.
\ee

\section{$\chi^2$ minimization}
We have used the SNe Ia data to constrain the parameters of
the variable Chaplygin gas model. Here, to find out the luminosity
distance we have taken into account the contributions from
radiation and baryons together with the Chaplygin gas.

Using the Friedmann equation [\ref{FRW}], in flat universe, the luminosity distance
can be expressed as

\be
d_L=\frac{c}{a H_0}\int_{a_{ls}}^1 \frac{da}{\Omega_{ch0}^{1/2} X(a)}
\ee

and the distance modulus

\be
\mu_{th}=5\log \frac{H_0 d_L}{c h} + 42.38
\ee

The best fit parameters are determined by minimizing ~\cite{wang}
\be
\chi^2=\sum_i\Big[\frac{\mu_{th}^i-\mu_{obs}^i}{\sigma_i}\Big]^2-\frac{C_1}{C_2}\Big(C_1+\frac{2}{5}\ln 10\Big)-2\ln h
\ee
where
\be
C_1 \equiv \sum_i \frac{[\mu_{th}^i-\mu_{obs}^i]}{\sigma_i^2}
\ee
\be
C_2 \equiv \sum_i \frac{1}{\sigma_i^2} 
\ee

We obtain the minimum $\chi^2=174.36$ for $\Omega_m=0.22$ and $n=-2.8 $ which are quite close
to those obtained by \cite{cgas}. The contour corresponding to $99\%~CL$ in
figure $1~\&~2$ gives the range of $\Omega_m=[0.0, 0.36]$ and $n=[-41.3, 2.8]$ as tabulated in Table 1.

\section{Discussion}
The Variable Chaplygin gas behaves like non-relativistic
matter during the early times  and has the ability
to drive the universe into an accelerated expansion in the recent
times. From SNe Ia Gold data, this model is consistent with wide range of
values for
parameters $\Omega_m$ and n. The best fit of the model obtained by Guo
\& Zhang \cite{cgas} gives $\Omega_m~=~0.25$ and $n~=~-2.9$ with
$\chi^2_{min}~=~173.88$. They performed the $\chi_{min}^{2}$ test  with the
gold data set \cite{reiss} using 
\be\label{FRW1}
H^2 = \frac{8\pi G}{3} \Big[\rho_{ch0} \Big[ \Omega_m(1+z)^6 +
    (1-\Omega_m)(1+z)^n \Big]^{1/2} \Big]
\ee
We repeated the test using the same SNe Ia Gold data set but took
into account contribution from matter as well as radiation, as given
in Eq. \ref{FRW}.  The results obtained by us give  $\chi^2_{min}~=~174.36$ at
$\Omega_m~=~0.22$ and $n~=~-2.8$. We find that  matter and radiation
do not contribute significantly to the expansion rate at redshifts for
which the supernova data is available but we include them for completeness
of our analysis. At $99\%$ confidence level the $\chi^{2}_{min}$ is
obtained for values of parameters  $\Omega_m$ and $n$ lying in the range [0.0,~0.36] and [-41.3,~2.8] respectively.  
\vskip 0.12 cm   

The Figures 1  and 2 show the 1$\sigma$, 2$\sigma$ and  3$\sigma$
$\chi^2$ contours of the SNe Ia Gold data and the location of CMBR
peaks (using results of WMAP) at 1$\sigma$ level in the Variable Chaplygin
Gas model. In Figure 1, the contours 1 and 2 represent, for the first
peak, the 1$\sigma$ lower ($l_{1}=291.3$) and the upper ($l_{1} = 220.9$) bounds on
$l$ obtained from WMAP data. In conclusion, the $1\sigma$ contours of the first peak of the WMAP
combined with SNe Ia Gold sample restrains the parameter space to
$\Omega_m = [0.017,~0.117]~~ \mathrm{and}~~ n = [-1.3,~2.6]$. In Figure
2, the contours 1 and 2 represent, for the third peak, the 1$\sigma$ lower ($l_{3}=802.0$) and the upper ($l_{3} = 816.0$) bounds on
$l$ obtained from WMAP data. The $1\sigma$ contours of the
third peak of the WMAP combined with SNe Ia Gold sample restricts
the parameter space to $\Omega_m = [0.018,~0.091]~\&~n =
[-0.2,~2.8]$. The results have been tabulated in Table 1.
\vskip 0.12 cm
In Figures  3 and 4 , we give the variation of the location of first
and third peak respectively with $n$ for different values of $\Omega_m$. The
horizontal lines in these two figures  show the observational $1 \sigma$
bounds for the two peaks. The bound on $\Omega_m $ from the
first peak is $\Omega_m \le 0.2$  and from the third peak is $\Omega_m
\le 0.18$.
\vskip 0.12 cm
Recently Guo and Zhang \cite{guo} have put constraints on the
Variable Chaplygin Gas using X-Ray gas mass fractions in galaxy
clusters and SNe Ia Gold data set. They have shown that the consistent
range of $\Omega_m$ and
$n$ are $ [0.043,~ 0.068]$ and $[1.18,~ 2.03]$ ($1 \sigma$ error
bar) respectively. Our results based on location of WMAP $ 3^{rd}$ peak and
SneIa at the $ 3 \sigma$ level indicate the range $[0.018,~
0.091]$ and $[-0.2,~ 2.8] $ for $\Omega_m$ and n respectively which
accommodates the results obtained from X-Ray gas mass fractions in
galaxy clusters and SNe Ia \cite{guo}. Though the original Chaplygin
Gas is  observationally ruled out, the Variable Chaplygin Gas can be  further explored.

\begin{center}
\bf Table 1. {\sl Bounds on $\Omega_m$ \& n.}
\end{center}
\begin{center}
\begin{tabular}{|c|c|c|c|}
\hline
\hline
\bf Method & \bf Reference & {\bf {$\Omega_m$}} & \bf n \\
\hline
\hline
 & & & \\
SNeIa($3\sigma$)& This paper & $\bf 0.22^{+0.14}_{-0.22}$ & $\bf -2.8^{+5.6}_{-38.5}$\\
 & & & \\
\hline
 & & & \\
WMAP $1^{st}$ Peak & This paper & \bf [0.017,~0.117] & \bf [-1.3,~2.6]\\
$+$ SNeIa($3\sigma$) & & & \\
 & & & \\
\hline
 & & & \\
WMAP $3^{rd}$ Peak & This paper & \bf [0.018,~0.091] & \bf [-0.2,~2.8]\\
$+$ SNeIa($3\sigma$) & & & \\
 & & & \\
\hline
\end{tabular}
\end{center}

\section*{Acknowledgments}

The authors thank Areejit Samal for useful discussions. Pranav Kumar
also acknowledges C.S.I.R., India for S.R.F. grant. The authors also
thank the referee for valuable comments and suggestions.



\pagebreak
\begin{figure}
\begin{center}
\epsfig{file=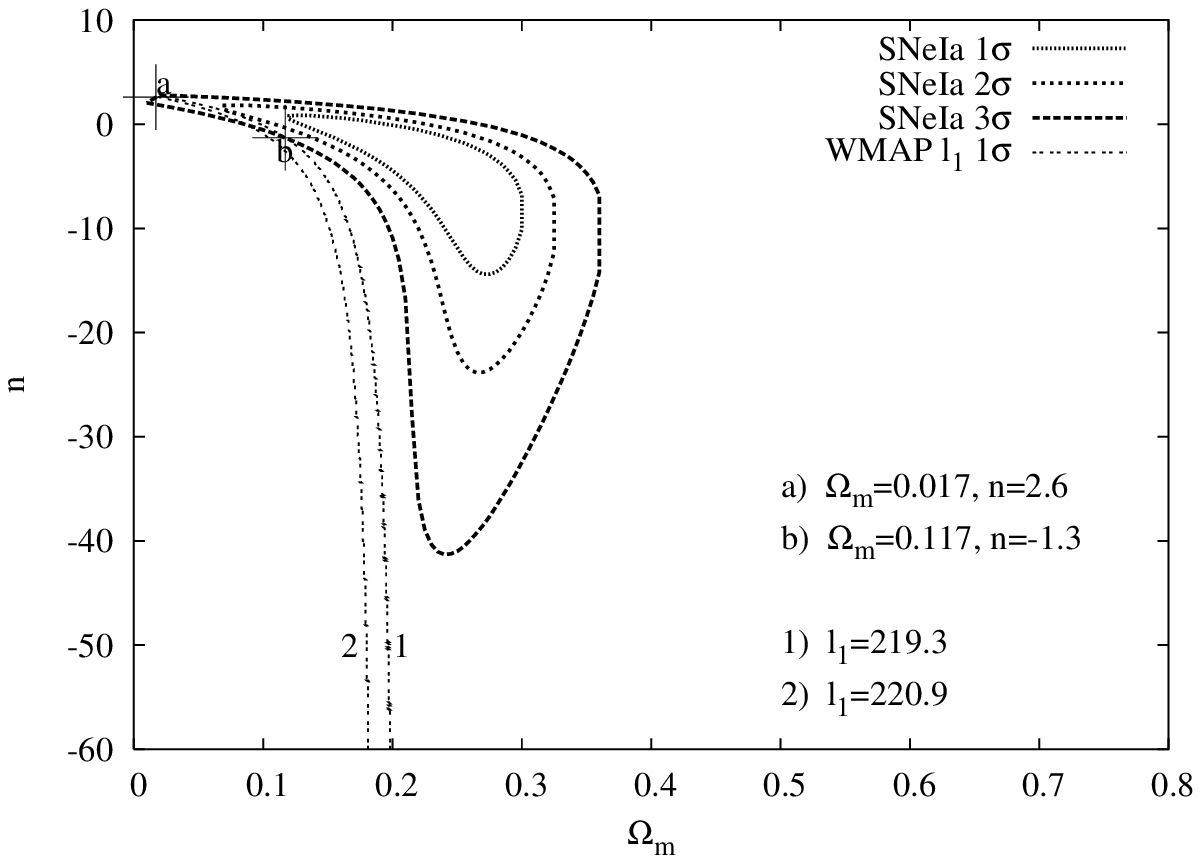,width=0.79\textwidth}
\caption{\it The combined WMAP $l_1$ and SNeIa parameter space}
\end{center}
\end{figure}
\begin{figure}
\begin{center}
\epsfig{file=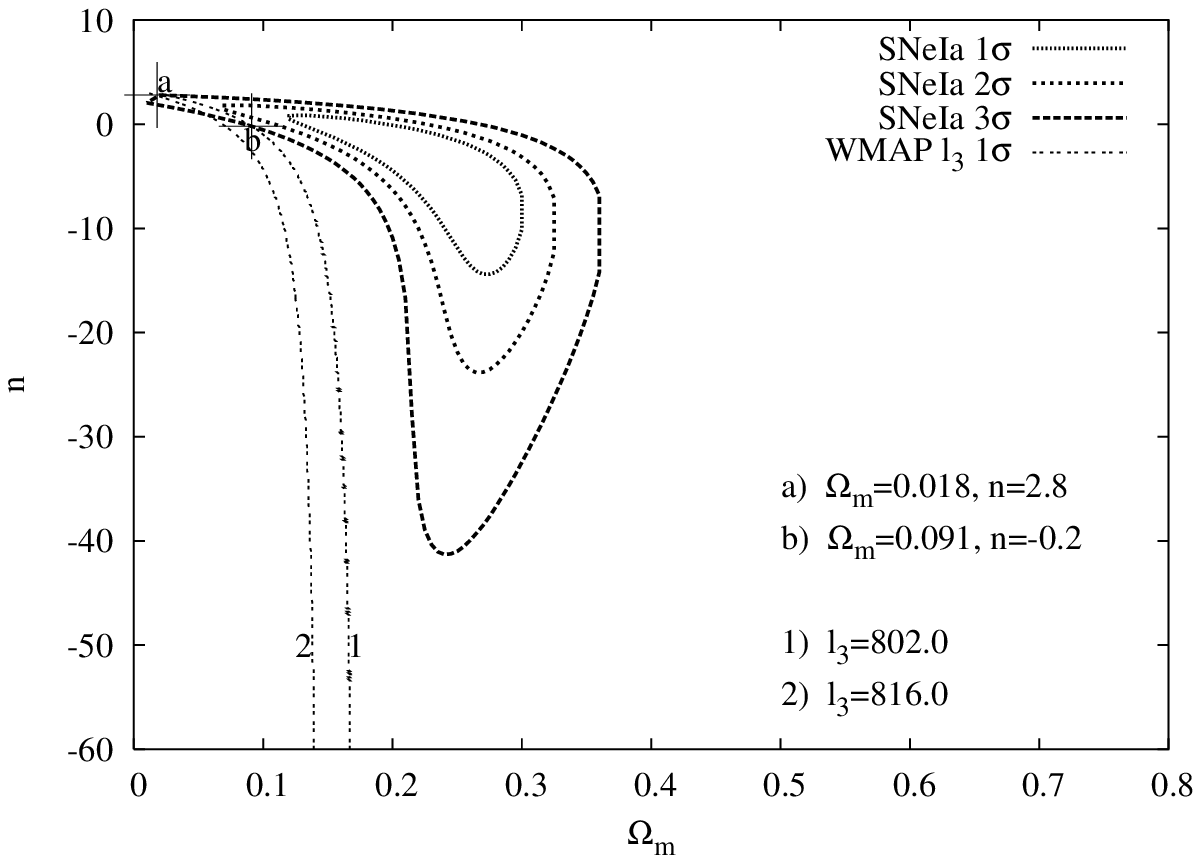,width=0.79\textwidth}
\caption{\it The combined WMAP $l_3$ and SNeIa parameter space}
\end{center}
\end{figure}
\pagebreak
\begin{figure}
\begin{center}
\epsfig{file=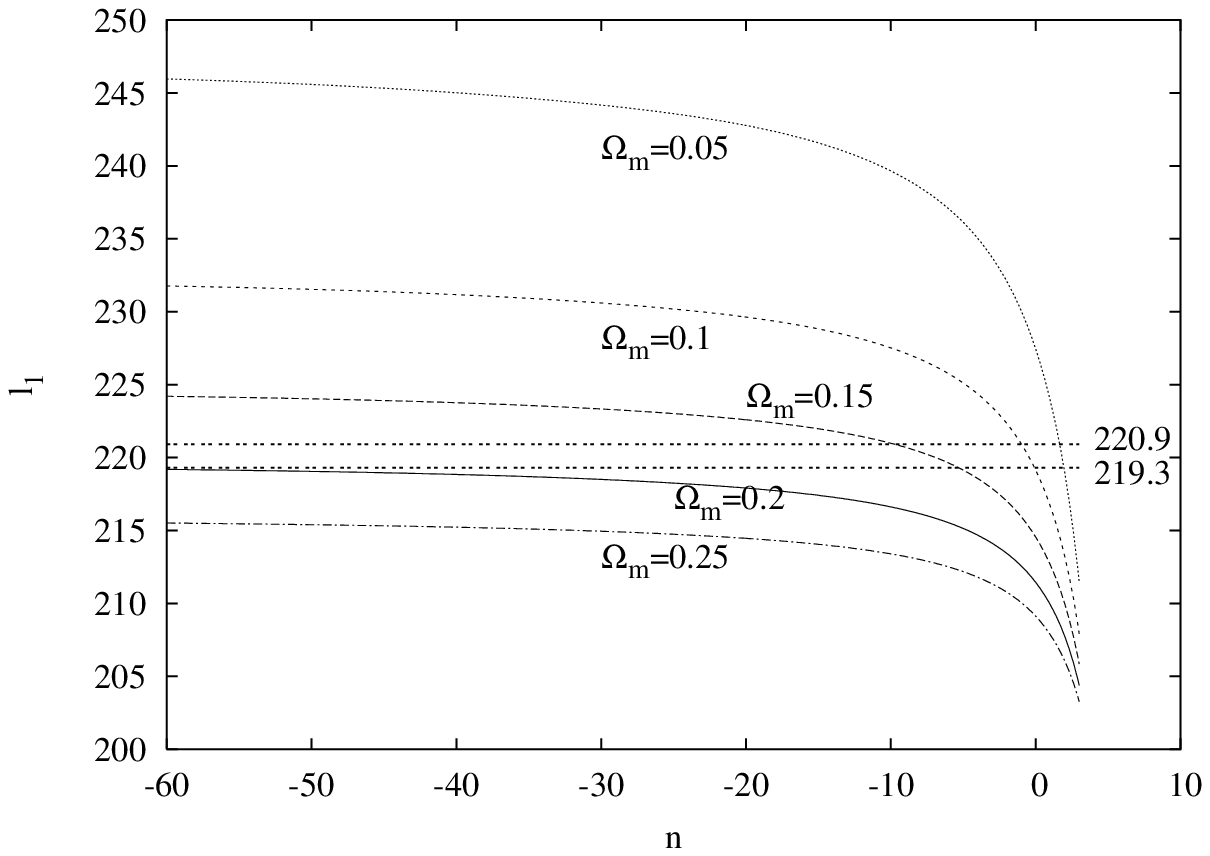,width=0.75\textwidth}
\caption{\it Variation of $l_1$ with  n for different $\Omega_m$. The
  horizontal lines correspond to the observational 1$\sigma$ bound on $l_1$ }
\end{center}
\end{figure}
\begin{figure}
\begin{center}
\epsfig{file=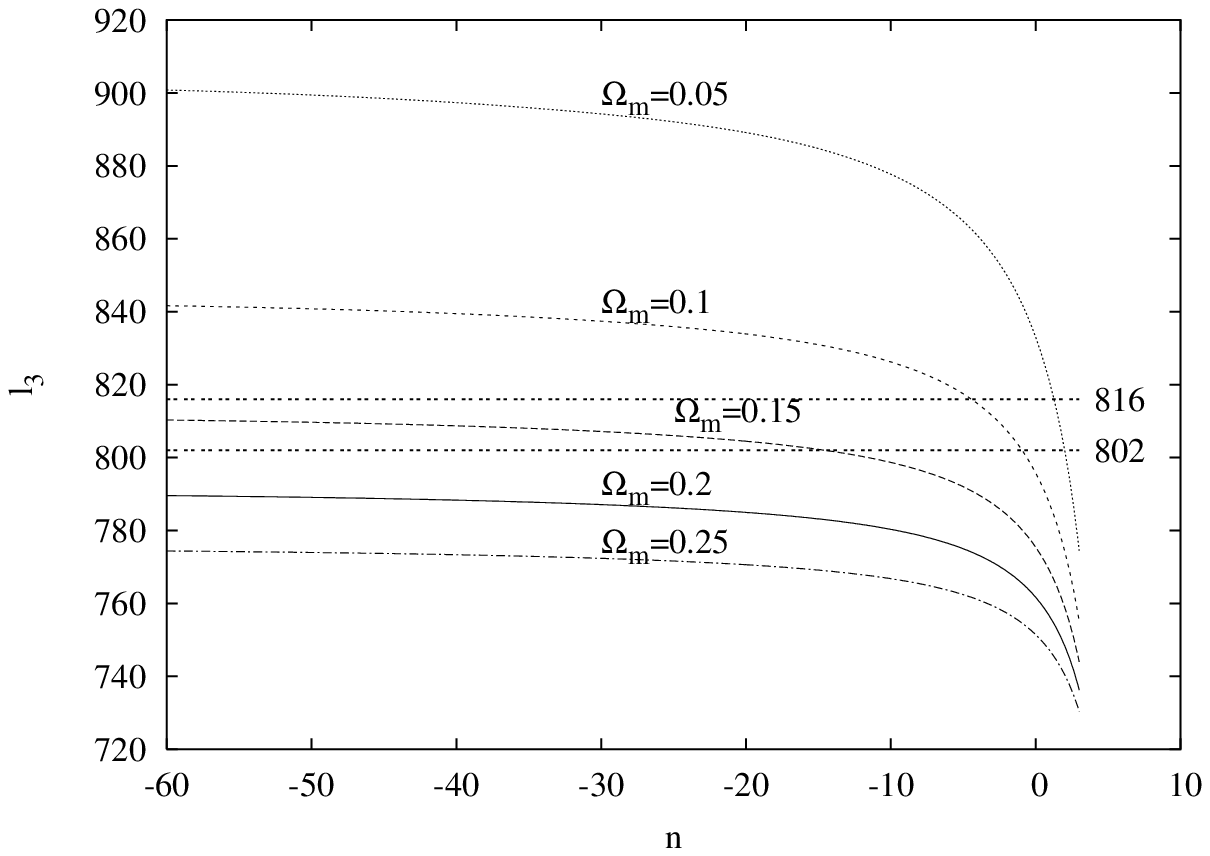,width=0.75\textwidth}
\caption{\it Variation of $l_3$ with  n for different $\Omega_m$. The
  horizontal lines correspond to the observational 1$\sigma$ bound on $l_3$ }
\end{center}
\end{figure}

\begin{thebibliography}{10}

\bibitem{perl} Perlmutter, S. {\it et al.}, {\it Nature}, {\bf 391}
  (1998), 51; Perlmutter, S. et al., \apj, {\bf 517} (1999), 565;
  Reiss, A.G.,{\it et al.}, {\it Astron.~J.}, {\bf 116} (1998), 1009.

\bibitem{varun} Peebles, P.J.E. \& Ratra, B., {\it
  Rev. Mod. Phys.}, {\bf 432} (2003), 559 ; Sahni, V. \& Starobainsky, A.,
  {\it Int.~J.~Modern~Phy. D}, {\bf9} (2000), 373; Padmanabhan, T.,
  {\it Phy. Rep.}, {\bf 380} (2003), 235.

\bibitem{weinberg} Weinberg, S. , {\it Rev.~Mod.~Phys.}, {\bf 61}  (1989), 1;

\bibitem{hkj} Jassal, H.K., Bagla, J.S. \& Padmanabhan, T., \mnras {\it Letters}, {\bf 356}
 (2005), L11-L16.

\bibitem{ratra} Ratra, B. \& Peebles, P.J.E., \prd, {\bf 37} (1988), 3406; Wetterich, C., {\it Nuc.~ Phy.}, {\bf B302} (1988), 668.

\bibitem{zlatev} Zlatev, I., Wang, L.M. \& Steinhardt, P.J., \prl, {\bf 82} (1999), 896; Steinhardt, P.J., Wang, L. \& Zlatev, I., \prd, {\bf 59} (1999), 123504; Lee, W. \& Ng K. W., \prd,  {\bf 67} (2003), 107302; Lee, D. S., Lee, W. \& Ng, K.W., {\it Int.J.Mod.Phys.},
{\bf D14} (2005), 335-344.

\bibitem{santos} Bento, M.C., Bertolami, O. \& Santos, N.C., \prd, {\bf 65} (2002), 067301.

\bibitem{kamen} Kamenshchik, A., Moschella, U. \& Pasquier, V., {\it Phy.~Lett.}, {\bf B511} (2001), 265; Gorini, V., Kamenshchik, A. \& Moschella, U., \prd, {\bf 67}, (2003), 063509.

\bibitem{billic}Bilic, N., Tupper, G.B. \& Viollier, R. D. ,{\it Phy.~Lett.}, {\bf B535} (2002), 17;

\bibitem{bento} Bento, M.C., Bertolami, O. \& Sen., A.A., \prd, {\bf 66} (2002), 43507.

\bibitem{makler} Makler, M., de~Oliveira, S.Q., \& Wang, I. \pl, {\bf
  B555} (2003), 1; Fabris, J.C., Goncalves, S.B.V. \& de~Souza, P.E.,
  {\bf astro-ph/0207430}; Gong, Y. \& Duan, C.K., {\it Class.~ Quant.~
    Grav.}, {\bf 21} (2004), 3655; \mnras, {\bf 352} (2004), 847;
  Gong, Y., {\it JCAP}, {\bf 0503} (2005), 007.


\bibitem{bento1} Bento, M.C., Bertolami, O. \& Sen., A.A., \prd, {\bf
  67} (2003), 063003; Carturan, D. \& Finelli, F., \prd, {\bf 68}
(2003), 103501.

\bibitem{dev} Dev, A., Alcaniz, J.S., Jain, D., \prd {\bf 67} (2003),
  023515;  Alcaniz, J. S., Jain, D.\& Dev, A.,\prd {\bf 67} (2003),043514;
 Zhu, Z.H., \AnA, {\bf 423} (2004), 421.

\bibitem{bean} Bean, R.\& Dore, O., \prd, {\bf 68} (2003), 023515;
Amendola, L., Finelli, F., Burigana, C. \& Carturan, D., {\it JCAP}, {\bf 0307} (2003), 005.

\bibitem{fabris} Fabris, J.C., Goncalves, S.B.V. \& de~Souza, P.E., \grg, {\bf 34} (2002), 53.

\bibitem{teg} Sandvik, H., Tegmark, M., Zaldarriaga, M. \& Waga, I., \prd
  {\bf 69} (2004), 123524.

\bibitem{cgas} Guo, Z.K. \& Zhang. Y.Z., {\bf astro-ph/0506091}.

\bibitem{reiss} Supernova Search Team Collaboration : Reiss, A.G. {\it et.al.},
  \apj, {\bf 607} (2004) 665.



\bibitem{fukugita} Hu, W., Fukugita, M., Zaldarriaga, M. \& Tegmark,
  M., {\it \apj}, {\bf 549}, (2001), 669.

\bibitem{doran} Doran, M. \& Lilly, M., {\it MNRAS}, {\bf 330},
  (2002), 965; Doran, M., Lilly, M., Schwindt, J. \& Wetterich, C.,
  {\it \apj}, {\bf 559}, (2001), 501.

\bibitem{wang} Wang, Y. et al., {\bf astro-ph/0402080}.
\bibitem{guo} Guo, Z.K. \& Zhang, Y.Z., {\bf astro-ph/0509790}.
\end{thebibliography}
\end{document}